%% file: non_markov_tf_refactor.tex
\documentclass[
twocolumn,groupedaddress,aps,prx,superscriptaddress,
amssymb,amsmath,amsfonts,10pt]
{revtex4-2}

\usepackage{bm}
\usepackage{graphicx}
\usepackage{subfigure}
\usepackage{algorithm}
\usepackage{algpseudocode}

\usepackage{wasysym} 

\begin{document}


\title{Quantify the Non-Markovian Process with Intervening Projections
in a Superconducting Processor}

\author{Liang Xiang}
\affiliation{Zhejiang Province Key Laboratory of Quantum Technology and Device,
Department of Physics, Zhejiang University, Hangzhou, 310027, China}

\author{Zhiwen Zong}
\affiliation{Zhejiang Province Key Laboratory of Quantum Technology and Device,
Department of Physics, Zhejiang University, Hangzhou, 310027, China}

\author{Ze Zhan}
\affiliation{Zhejiang Province Key Laboratory of Quantum Technology and Device,
Department of Physics, Zhejiang University, Hangzhou, 310027, China}

\author{Ying Fei}
\affiliation{Zhejiang Province Key Laboratory of Quantum Technology and Device,
Department of Physics, Zhejiang University, Hangzhou, 310027, China}

\author{Chongxin Run}
\affiliation{Zhejiang Province Key Laboratory of Quantum Technology and Device,
Department of Physics, Zhejiang University, Hangzhou, 310027, China}

\author{Yaozu Wu}
\affiliation{Zhejiang Province Key Laboratory of Quantum Technology and Device,
Department of Physics, Zhejiang University, Hangzhou, 310027, China}

\author{Wenyan Jin}
\affiliation{Zhejiang Province Key Laboratory of Quantum Technology and Device,
Department of Physics, Zhejiang University, Hangzhou, 310027, China}

\author{Zhilong Jia}
\affiliation{Key Laboratory of Quantum Information, University of Science and
Technology of China, Hefei, 230026, China}

\author{Peng Duan}
\affiliation{Key Laboratory of Quantum Information, University of Science and
Technology of China, Hefei, 230026, China}

\author{Jianlan Wu}
\affiliation{Zhejiang Province Key Laboratory of Quantum Technology and Device,
 Department of Physics, Zhejiang University, Hangzhou, 310027, China}

\author{Yi Yin}
\email{yiyin@zju.edu.cn}
\affiliation{Zhejiang Province Key Laboratory of Quantum Technology and Device,
Department of Physics, Zhejiang University, Hangzhou, 310027, China}

\author{Guoping Guo}
\email{gpguo@ustc.edu.cn}
\affiliation{Key Laboratory of Quantum Information, University of Science and
Technology of China, Hefei, 230026, China}
\affiliation{Origin Quantum Computing, Hefei, 230026, China}


\begin{abstract}
A Markov assumption considers a physical system memoryless to simplify its dynamics.
Whereas memory effect or the non-Markovian phenomenon is more general in nature.
In the quantum regime, it is challenging to define or quantify the non-Markovianity
because the measurement of a quantum system often interferes with it.
We simulate the open quantum dynamics in a superconducting processor,
then characterize and quantify the non-Markovian process. With the complete set of
intervening projections and the final measurement of the qubit, a restricted process
tensor can be determined to account for the qubit-environment interaction.
We apply the process tensor to predict the quantum state with memory effect,
yielding an average fidelity of $99.86\%\pm 1.1\permil$. We further derive the Choi
state of the rest process conditioned on history operations and quantify the
non-Markovianity with a clear operational interpretation.
\end{abstract}


\maketitle


\section{Introduction\label{sec_I}}

Schrödinger equation can formulate the evolution of a quantum system whose
state is isolated.
A realistic quantum system is unavoidably coupled to environment,
which typically contains many degrees of freedom and hard to characterize and
control~\cite{ChuangBook,Breuer2002OQDbook,Rivas2012OQSBook}.
The dynamics of this open quantum system sometimes can be simplified with a
Markov assumption, such as in the Gorini-Kossakowski-Sudarshan-Lindblad (GKSL)
master equation~\cite{KossakowskiRMPhys1972,GKSJMP1976,LindbladCMPhys1976}.
It assumes that the environment is stationary and
the system-environment interaction is weak and uncorrelated.
Mathematically, a Markov evolution generated by the GKSL master equation is
composed of a series of completely positive trace-preserving (CPTP) maps,
which form a semi-group~\cite{GKSJMP1976,LidarOQS2019}.

If the evolution of a quantum system depends not only on its instantaneous state but also on history, it is considered non-Markovian or has memory effects.
The information of the process history can be
stored somewhere and retrieved later to affect the system.
The non-Markovian behavior in the quantum regime is challenging to define and quantify,
mostly because that general measurements of a quantum system will unavoidably
interfere with it~\cite{RivasRPPhys2014,PollockPRL2018}.
With the advance of quantum device and control apparatus, the revealing of the
non-Markovian effects have been constantly reported
~\cite{MalleyPRApp2015,
RingbauerRPL2015,YuPRL2018,LuPRL2020,ChenPRApplied2020,SamachArXiv2021}.
Previous studies of non-Markovianity mainly characterize the
open dynamics with the final tomographic measurement, i.e. the observer only
waits at the end of the stochastic process to measure the mixed quantum state,
and the non-Markovianity is revealed by examining the state changes over a
series of time steps.
In general, there are two kinds of approaches to characterizing the
non-Markovianity ~\cite{BreuerPRL2009,RivasPRL2010,
RingbauerRPL2015,YuPRL2018,LuPRL2020,
BreuerRevModPhys2016,LiPhysRep2018,LiEPL2019,LiEPL2020}.
One is to detect the information flow between the system and its
environment. The other is to check the divisibility of the dynamical map.

Although previous experiments have shown evidence of the quantum non-Markovian
process, there is no consensus on the quantification of memory
effect~\cite{PollockPRL2018,MilzPRL2019}.
The observations also lacks a clear causal
structure~\cite{CostaNJP2016,MilzQuantum2020}.
Here in this work, we apply an alternative method to characterize and quantify
the non-Markovian process with intervening projective measurements,
or positive operator-valued measurements (POVMs)~\cite{ChuangBook}.
The projections are not only the final measurement but also active
interrogations on each intermediate step of the process.
This method is feasible because that projective measurements can be
integrated as local operators in the tomographic process tensor framework
~\cite{PollockPRA2018,MilzPRA2018}.
Experimentally, we design a three-time-step open quantum process in a
superconducting processor, where standard two-qubit gates are selected to
simulate the system-environment interaction.
We implement arbitrary POVMs by using the field-programmable-gate-array
(FPGA)-based fast measurement and control hardware and customized quantum
instruction set architecture (ISA)~\cite{XiangPRAppl2020}.
The restricted process tensor is determined using a complete set of POVMs.
The intervening projective measurements reveal the information of the
instantaneous state, and at the same time refresh the system deterministically,
from which we can compare the causal relation on different time steps.

Note that a process tensor has been experimentally determined with unitary gates
on IBM's cloud-based quantum processors~\cite{WhiteNatComm2020}.
Their work traces the information flow between different stages of the process,
and provides the lower bounding of the memory effect. Here since projective
measurements destroy the system-environment entanglement and steer the system
into a definite pure state that is independent of its previous trajectories,
we can check if there is a conditional dependence of the future dynamics on
the past control operations~\cite{PollockPRL2018,MilzPRA2018}.
Consequently, we can directly determine whether a process is Markovian or
non-Markovian in a finite number of experiments.
With projective measurements providing complete information of the system
during the process, we finally quantify the non-Markovian process in a
smaller subset of time steps conditioned on earlier operations.
Based on our experiment, we illustrate the operational interpretation of
the non-Markovianity.

\section{Open Quantum Dynamics Described by the Tensor Network}

\begin{figure}[htp]
\centering
\includegraphics[width=1.0\columnwidth]{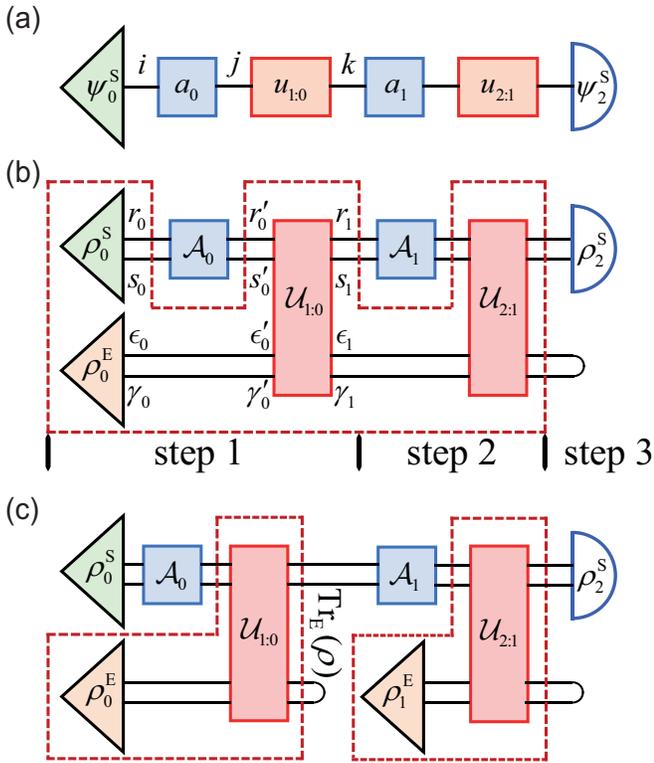}
\caption{
Tensor diagrams of the three-time-step process,
including initial states of the system $\mathrm{\mathbf{S}}$ (green triangle)
and the environment $\mathrm{\mathbf{E}}$ (orange triangle),
intervening local operators (blue box)
and $\mathrm{\mathbf{SE}}$ interactions (red rectangle).
(a) Closed dynamics of a single qubit $\mathrm{\mathbf{S}}$.
(b) Process tensor depiction of the open quantum dynamics.
The process tensor $\mathbf{\mathcal{T}}^{2:0}$ encodes
all the environmental information in the dashed frame.
(c) Traditional depiction of the open quantum dynamics, represented by
a concatenation of CPTP maps. For each step, the dynamics of
$\mathrm{\mathbf{S}}$ is simulated by the reduced $\mathrm{\mathbf{SE}}$
unitary evolution (outlined in the dashed frame).
}
\label{fig01}
\end{figure}

We here use the mathematical tool of
tensor network~\cite{PenroseThesis1956,PenroseCMAppPenrose1971}
to describe the quantum process.
In a simplified situation [Fig.~\ref{fig01}(a)], a complex vector
describes the pure state of a qubit system $\mathrm{\mathbf{S}}$.
It can be equivalently denoted by a tensor $(\psi_0)_i$ with one index
$\textit{i}=$0 or 1, encoding the ground or excited state.
At each time step $n$, the system evolution can
be represented by a tensor $(\mathbf{u}_{n:n-1})^j_k$,
while an observer intervenes in the trajectory with a local operator
$(\mathbf{a}_{n-1})^i_j$. The dynamics of
$\mathrm{\mathbf{S}}$ can be simulated by summing over the
corresponding tensor indexes,
called a `tensor contraction'~\cite{BiamonteArXiv2017}.

For an open quantum system, we design a process where a single qubit
$\mathrm{\mathbf{S}}$ interacts with its environment [Fig.~\ref{fig01}(b)].
The state of $\mathrm{\mathbf{S}}$ under stochastic quantum evolution is now
represented by a density matrix $\rho^{\mathrm{S}}_{s_0,r_0}$.
In our superconducting processor, we choose a neighboring ancilla qubit
$\mathrm{\mathbf{E}}$ to simulate the environment of $\mathrm{\mathbf{S}}$.
Initially, $\mathrm{\mathbf{S}}$ and $\mathrm{\mathbf{E}}$
are prepared to a tensor product of the ground states,
$\rho^{\mathrm{SE}}_{s_0,r_0;\gamma_0,\epsilon_0}
=\rho^{\mathrm{S}}_{s_0,r_0}
\otimes
\rho^{\mathrm{E}}_{\gamma_0,\epsilon_0}
=|00\rangle$.
The intervened local operator on $\mathrm{\mathbf{S}}$ is represented by a
tensor $\mathcal{A}^{s_0,r_0}_{s'_0,r'_0}$, and
the $\mathrm{\mathbf{SE}}$ interaction is represented by a tensor
$\mathcal{U}^{s'_0,r'_0;\gamma'_0,\epsilon'_0}_{s_1,r_1;\gamma_1,\epsilon_1}$.
This $n$-step quantum process can be simulated by contracting
$\rho^{\mathrm{\mathbf{SE}}}$ with sequential $\mathcal{A}$s and
$\mathcal{U}$s as following,
\begin{align}
\rho_{n}{(\mathbf{A}_{n-1: 0})}
& = \operatorname{Tr}_\mathrm{E}
\left[
\mathcal{U}_{n: n-1} \mathcal{A}_{n-1}
\cdots
\mathcal{U}_{1: 0} \mathcal{A}_{0}
\left
(\rho_{0}^\mathrm{SE}\right)
\right]
\label{eq_tensor_contract} \\
& = \mathcal{T}^{n: 0}\left[\mathbf{A}_{n-1: 0}\right].
\label{eq_process_tensor}
\end{align}
After tracing out the environment indexes at the final step
[Eq.~(\ref{eq_tensor_contract})],
we derive the output state of $\mathrm{\mathbf{S}}$.
Consequently, there is a map from the sequence of local operators
$\{\mathcal{A}_0,\cdots,\mathcal{A}_{n-1}\}$ ($\mathbf{A}_{n-1:0}$ for short)
to the output state $\rho_{n}$.
This map is coined a `quantum comb' in the quantum circuit
architecture~\cite{ChiribellaPRL2008},
or a `process tensor' in open quantum dynamics~\cite{PollockPRA2018}.

The process tensor encodes the hidden environmental information in
the dashed frame of Fig.~\ref{fig01}(b).
In Eq.~(\ref{eq_process_tensor}), $\mathbf{\mathcal{T}}^{n:0}$ is a
multilinear map on $\mathbf{A}_{n-1:0}$, i.e.
its linearity holds independently to the operator $\mathcal{A}$ at each step.
Then $\mathcal{T}^{n: 0}$ can be experimentally determined by the quantum
tomography technique~\cite{FanoRMP1957}.
At step $n$, the local operator can be uniquely decomposed by a fixed set
of linearly independent basis operators $\{\mathcal{B}_l\}$,
$\mathcal{A}_{n-1}=\sum_{l}(\beta_{n-1})_l\mathcal{B}_l$.
The sequence of local operations can be further expanded in terms of
their tensor products,
\begin{align}
\mathbf{A}_{n-1: 0}=\sum_{\vec{l}} \beta_{\vec{l}}
\bigotimes_{m=0}^{n-1}
(\mathcal{B}_{m})_l
\label{eq_ops_A_of_B}
\end{align}
where real numbers $\beta_{\vec{l}}$ are the coefficients of the tensor products
of $\mathcal{B}s$, and $\vec{l}:=\{l_{0}, l_{1}, \cdots, l_{n-1}\}$
denotes the combination of operators.

A process tensor has been recently determined using a set of unitary control
gates. The tensor can be applied to characterize a non-Markovian
process~\cite{WhiteNatComm2020}.
In this work, we probe the system with intervening projective measurements
or POVM operators.
Experimentally, our FPGA-based hardware and customized quantum ISA~\cite{XiangPRAppl2020}
enable us to apply an arbitrary POVM operator on any
time step of the process.
These POVMs effectively destroy the entanglement between the system and the
environment, and steer the system to a definite pure state independent
of its previous trajectories~\cite{PollockPRL2018}.
With a complete set of POVM operators $\{\mathcal{P}\}$, a `restricted'
process tensor $\mathbf{\mathcal{T}}_{P}$ can be determined~\cite{KuahPRA2007}.
We then derive the process tensor in a subset of steps, which is `contained'
in a larger one (the containment property)~\cite{PollockPRA2018}.
With the smaller process tensor, we quantify the system's non-Markovianity by
calculating the von Neumann mutual information~\cite{VedralRMP2002}
of its Choi state~\cite{ChoiLAA1975,ChoiJisomorphism},
and give a clear operational interpretation in the context of our experiment.


\section{Results}

In our experiment, the superconducting quantum processor is a chip composed of
six cross-shaped Transmon qubits~\cite{WangPRAppl2019}.
The device was mounted in a dilution refrigerator whose base temperature is
around 10 mK.
We choose two neighboring qubits as the system $\mathrm{\mathbf{S}}$ and
the environment $\mathrm{\mathbf{E}}$.
They are set at $\omega_\mathrm{S}=6.21$ GHz and $\omega_\mathrm{E}=5.70$ GHz.
The other device parameters are the same as that in Ref.~\cite{ZhanArXiv2021}.
We specifically design two distinct processes consisting of standard quantum
gates in the quantum processor.
One has the $\mathrm{\mathbf{SE}}$ interaction brought by a CNOT gate after the
first operator and CZ gate the second. We swap these
two gates for the other process, where CZ is activated first and CNOT the second.

\subsubsection{Implementation of POVMs}
\label{sec_povm}

The sub-normalized state of the quantum system after a POVM operator
$\mathcal{P}$ is given by $\mathcal{P} \rho \mathcal{P}^{\dagger}$,
where $\mathcal{P}=|p\rangle\langle p|$ also denotes the state to which we
project. The probability of this projection is
$\operatorname{Tr}[\mathcal{P} \rho]$.
An important consequence of the POVM operator on an open quantum system is
the destruction of its entanglement with the environment, as
$(\mathcal{P}^\mathrm{\mathbf{S}}\otimes\mathcal{I}^\mathrm{\mathbf{E}})
\rho^\mathrm{\mathbf{SE}}
(\mathcal{P}^\mathrm{\mathbf{S}}\otimes\mathcal{I}^\mathrm{\mathbf{E}})
=\mathcal{P}^\mathrm{\mathbf{S}}\otimes\mathcal{I}^\mathrm{\mathbf{E}}
\operatorname{Tr}_\mathrm{\mathbf{S}}
[(\mathcal{P}^\mathrm{\mathbf{S}}\otimes\mathcal{I}^\mathrm{\mathbf{E}})
\rho^\mathrm{\mathbf{SE}}]$, where $\mathcal{I}^\mathrm{\mathbf{E}}$ is the
identity operator on the environment state.

\begin{figure}[htp]
\centering
\includegraphics[width=1.0\columnwidth]{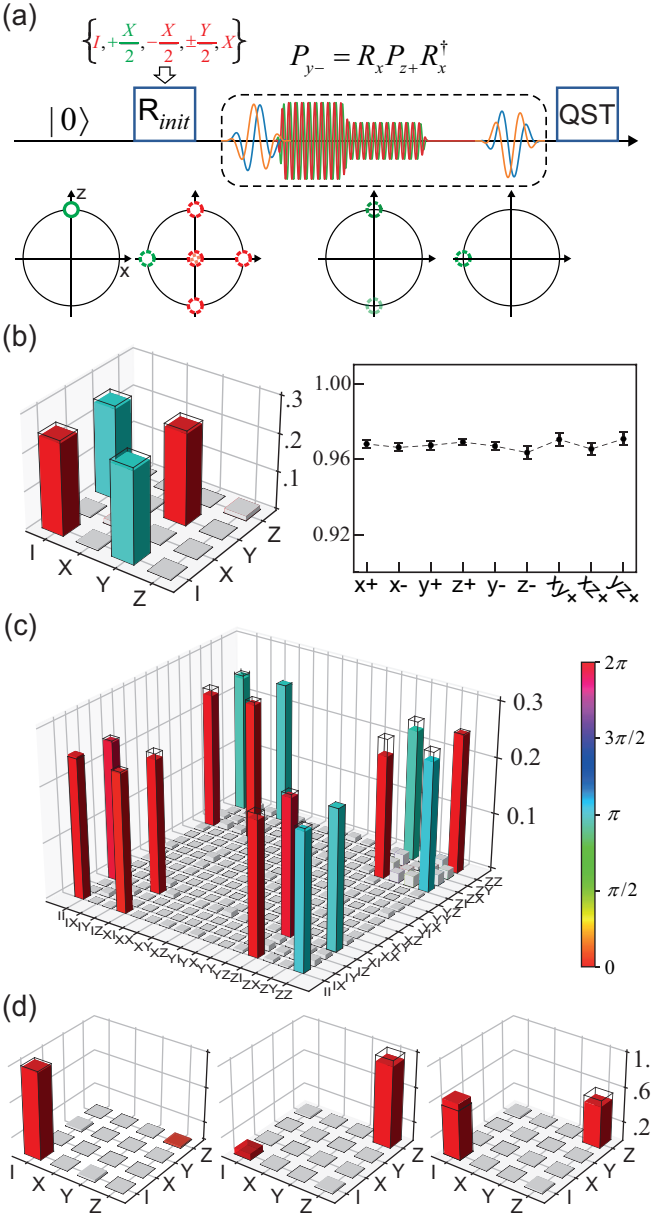}
\caption{
(a) QPT protocol to characterize the $\chi$-matrix of $\mathrm{\mathbf{S}}$,
with the implementation of $\mathcal{P}_{y-}$ as an example.
(b) Left panel: $\chi$-matrix of $\mathcal{P}_{y-}$.
The magnitude and complex phase of $\chi$ elements are shown as the height
and color of cubic columns. Magnitudes lower than 0.02 are drawn in gray
for visibility. Right panel: QPT fidelities of nine POVMs.
(c) $\chi$-matrix of the two-qubit CZ gate.
(d) $\chi$-matrices of the reduced CZ-gate, with $\mathrm{\mathbf{E}}$
being prepared to the ground state (left), excited state (middle), and
$(|0\rangle-i|1\rangle)/\sqrt{2}$ (right).
}
\label{fig02}
\end{figure}

The qubit is usually read out in a superconducting processor by projecting it to
the ground state ($\mathcal{P}_{z+}=|0\rangle\langle0|$) or excited state
($\mathcal{P}_{z-}=|1\rangle\langle1|$).
Whereas in general cases, $\mathcal{P}$ shall project the qubit to any state on
the Bloch sphere, which is parameterized as
$|p\rangle=\mathrm{cos}\frac{\theta}{2}|0\rangle
+ e^{i\phi}\mathrm{sin}\frac{\theta}{2}|1\rangle$.
Any $\mathcal{P}(\theta, \phi)$ can be realized by introducing the following unitary transformation,
\begin{align}
\mathcal{P}\rho \mathcal{P}^{\dagger}
&=
(R \mathcal{P}_{z+} R^{\dagger}) \rho
(R \mathcal{P}_{z+} R^{\dagger})^{\dagger}\notag \\
&=
R(\mathcal{P}_{z+}\operatorname{Tr}[\mathcal{P}_{z+} R^{\dagger} \rho R])
R^{\dagger}
\label{eq_povm_unitary_T}
\end{align}
where $R(\theta, \phi)$ is a $\theta$-angle rotation around the vector
$\vec{n}$ in the $xy$-plane. The azimuth of $\vec{n}$ is $\phi+\pi/2$.
As an example, the experimental sequence to implement a POVM operator
$\mathcal{P}_{y-}$ is drawn in the dashed frame in Fig.~\ref{fig02}(a).
We first rotate the qubit by $+\pi/2$ around the $x$-axis.
Then we apply a fast dispersive measurement~\cite{JeffreySankPRL2014}
to project it to either the ground state $\mathcal{P}_{z+}$
(the upper ball)
or excited state $\mathcal{P}_{z-}$ (the lower ball with lighter color).
Finally, we rotate the qubit back to the target state
$|p\rangle$ (ball on the rightmost Bloch sphere along the -$y$-axis).
Note that in Eq.~\ref{eq_povm_unitary_T} the probability of $\mathcal{P}$ is
calculated by post-selecting~\cite{JohnsonPRL2012} the ground-state
($\mathcal{P}_{z+}$).
Similarly, we can implement other POVMs to project the qubit onto any
axis.
$\mathcal{P}_{x+}$ denotes the projection on the positive $x$-axis;
$\mathcal{P}_{yz+}$ or $\mathcal{P}_{zy+}$ denotes the projection on the
internal ($\theta=-\pi/4$, $\phi=0$) or exterior ($\theta=\pi/4$, $\phi=0$)
angle bisector of +$y$ and +$z$ axis; $\mathcal{P}_{yz-}$ is the
projection on the reversed direction of $\mathcal{P}_{yz+}$, and so on.

\subsubsection{Traditional Picture of the Quantum Map}
\label{sec_cptp_result}

The CPTP map of each step of the process can be equivalently expressed in terms
of the Kraus decomposition~\cite{ChuangBook,VedralRMP2002,MilzOSID2017},
\begin{align}
\Lambda(\rho)=\sum E_{m} \rho E_{n}^{\dagger} \chi_{m n}
\label{eq_kraus_chi}
\end{align}
where the set of matrices $\{E\}$ are called the Kraus operators of $\Lambda$.
To benchmark the quantum gates, QPT protocol are normally used to determine the
matrix $\chi_{m n}$~\cite{ChuangBook,YamamotoPRB2010}.
Here we choose Pauli operators $\{I, \sigma_x, \sigma_y, \sigma_z\}$ as the
Kraus operators.
The gate fidelity is calculated by
$\operatorname{Tr}[\chi_\mathrm{ideal} \chi]$, where $\chi_\mathrm{ideal}$
represents an ideal matrix.

We apply the QPT protocol to characterize the POVMs mentioned in
Sec.~\ref{sec_povm}.
In our experiment, we prepare six qubit states by rotations on the ground state
$|0\rangle$.
After the process [dashed frame of Fig.~\ref{fig02}(a)], the output states are
determined using quantum state tomography (QST)~\cite{ChuangBook,SteffenPRL2006}.
Through out the paper, we use six symmetric POVMs
($\mathcal{P}\{$$x\pm$,$y\pm$,$z\pm\}$) to determine the quantum state,
though four informationally complete (IC) POVMs are sufficient~\cite{FlammiaFoundPhys2005}.
Knowing the input-output relation on a complete set of basis, it is sufficient
to determine $\chi$.
We can see four bars standing symmetrically on both side of the diagonal of the
matrix [left panel of Fig.~\ref{fig02}(b)], representing
coefficients of[1/4, -1/4, -1/4, 1/4] for Kraus operators
[$(I, I)$, $(I, \sigma_y)$, $(\sigma_y, I)$, $(\sigma_y, \sigma_y)$].
Not surprisingly, the quantum map we determine coincides with
$\mathcal{P}_{y-}$,
\begin{align}
\Lambda_{\mathcal{P}_{y-}}[\rho]
&\approx
\frac{1}{4}(I \rho I - \sigma_y \rho I - I \rho \sigma_y + \sigma_y \rho \sigma_y)
\notag \\
&=\frac{I - \sigma_y}{2} \rho \frac{I - \sigma_y}{2}
=\mathcal{P}_{y-} \rho \mathcal{P}_{y-}.  \notag
\end{align}
Note that while the projective measurement is performed along a
deterministic axis, the two complementary POVM elements it
contains is probabilistic and may not necessarily be trace preserving (TP).

We also characterize other POVMs, which will be used in Sec.~\ref{sec_ptensors}
as a full basis of operators to determine the process tensor.
After normalization, the fidelity of $\mathcal{P}_{y-}$ is 96.9\% $\pm$0.12\%,
averaged over 20 independent QPTs.
Fidelities of other POVMs are plotted in the right panel of Fig.~\ref{fig02}(b).
We have confirmed that the POVM operators are achieved with high fidelities
before intervening them in a quantum process to determine the process tensor.

Similarly, we characterize the $\chi$ matrices of the CZ gate
[Fig.~\ref{fig02}(c)] and CNOT gate (not shown).
Their fidelities are $96.3\%\pm0.29\%$ and $93.1\%\pm0.53\%$, respectively.
We next characterize the quantum map of $\mathrm{\mathbf{S}}$ when it interacts
with the ancilla $\mathrm{\mathbf{E}}$, termed the `reduced' quantum map of
$\mathrm{\mathbf{S}}$.
For CZ gate, the `reduced' map of $\mathrm{\mathbf{S}}$ depends on the
state of $\mathrm{\mathbf{E}}$, which can be written as,
\begin{align}
\Lambda_{\operatorname{Tr}_\mathrm{\mathbf{E}}[\mathrm{CZ}]}
(\rho^\mathrm{\mathbf{S}})=
\operatorname{Tr}_\mathrm{\mathbf{E}}[\mathcal{U}_\mathrm{CZ}
\rho^\mathrm{\mathbf{SE}} \mathcal{U}^\dagger_\mathrm{CZ}].
\label{eq_chi_reduced_cz}
\end{align}

Figure~\ref{fig02}(d) displays the $\chi$-matrices of three
`reduced' CZ gates with $\mathrm{\mathbf{E}}$ in
the ground state ($|0\rangle$),
the excited state ($|1\rangle$),
and the superposition state ($(|0\rangle-i|1\rangle)/\sqrt{2}$).
The equivalent quantum maps are
an identity ($I$),
a $180^\circ$ rotation around the $z$-axis ($Z$),
and a complete phase erasing (mixture of $I$ and $Z$ with equal probability), respectively.

\subsubsection{Tomography of the Restricted Process Tensor}
\label{sec_ptensors}

Following we choose a complete set of basis POVM operators~\cite{KuahPRA2007},
\begin{align}
\mathcal{F}=\mathcal{P}{\{x+, x-, y+, z+, y-, z-, xy+, xz+, yz+\}}.
\end{align}

By performing the combinations
$\{\mathrm{\mathbf{B}}_{1:0}\}$ = $\bigotimes^{2}\{\mathcal{P}\}$ at the first
two steps of the process and measuring the outcome states of $\mathrm{\mathbf{S}}$
using QST, the restricted process tensor can be determined as the following
algorithm,
\begin{algorithm}[H]
\renewcommand{\thealgorithm}{}
\caption{Process Tomography of $\mathbf{\mathcal{T}}_{P}^{2:0}$}
\begin{algorithmic}[1]
\State Stats = 3000
\Comment{Repetitions of the sequence}
\State $\mathrm{\mathbf{U}}_{2:0}$ = CNOT-CZ or CZ-CNOT
\Comment{$\mathrm{\mathbf{SE}}$ interactions}
\State $\{\mathrm{\mathbf{B}}_{1:0}\}=
\bigotimes^{2}\{P_{x+},P_{x-},P_{y+},\ldots,P_{yz+}\}_{9\times9}$
\Comment{three-step Basis}
\For {$\mathcal{B}_{l_{0}},\mathcal{B}_{l_{1}}$ in $\{\mathrm{\mathbf{B}}_{1:0}\}$}
    \For {$R_{m}$ in \{I, X/2, Y/2, -X/2, -Y/2, X\}}
    \Comment{QST}
        \For {i in range(Stats)}
            \State Apply $\mathcal{B}_{l_{0}}$ to $\mathrm{\mathbf{S}}$
            \State Apply $\mathcal{U}_0$ to $\mathrm{\mathbf{SE}}$
            \Comment{1st $\mathrm{\mathbf{SE}}$ interaction}
            \State Apply $\mathcal{B}_{l_{1}}$ to $\mathrm{\mathbf{S}}$
            \State Apply $\mathcal{U}_1$ to $\mathrm{\mathbf{SE}}$
            \Comment{2nd $\mathrm{\mathbf{SE}}$ interaction}
            \State Apply $\mathrm{R}_{m}$ to $\mathrm{\mathbf{S}}$
            \State Record the system state of this sequence $S_{\vec{l},m,i}$
        \EndFor
        \State {Compute the state probability $P_{\vec{l},m}$
                from $\{S_{\vec{l},m,i}\}$}
    \EndFor
    \State {Compute the QST result $(\rho_{\vec{l}})_{2\times2}$
            from $\{P_{\vec{l},m}\}$}
\EndFor
\State {Find the solution
        $(\mathbf{\mathcal{T}}_{P}^{2:0})_{256\times4}$
        as a linear map from
        $(\mathrm{\mathbf{B}}_{{\vec{l}}})_{9\cdot9\times4^2\cdot4^2}$
        to
        $(\rho_{{\vec{l}}})_{9\cdot9\times2\cdot2}$}
\end{algorithmic}
Note we fit the restricted process tensor using the least-square method
provided by scientific computing packages~\cite{HarrisNumpyNat2020},
and then obtain the best approximation $\mathbf{\mathcal{T}}_{P}$
with positive semi-define constraint and convex optimization
(see Sec.~\ref{quantify_NM}).
\end{algorithm}
\label{alg_1}

Once a process tensor $\mathbf{\mathcal{T}}_{P}^{2:0}$ is determined,
we can predict its output state when arbitrary projections
$\mathrm{\mathbf{A}}_{1:0}$ are inserted in the process.
The fidelity of predicted state is defined as~\cite{JozsaJModOpt1994},
\begin{align}
F(\rho, \sigma)=
(\operatorname{Tr} \sqrt{\sqrt{\rho} \sigma \sqrt{\rho}})^{2}
\label{eq_fidelity}
\end{align}
where $\rho$ is the density matrix of $\mathrm{\mathbf{S}}$ measured in
experiment, and $\sigma$ is the predicted one.
We use an over complete set of $\mathrm{\mathbf{A}}_{1:0}$,
consisting of POVMs in
$\mathcal{P}\{x+$, $x-$, $y+$, $z+$, $y-$, $z-$, $xy+$, $xz+$, $yz+$,
$xy-$, $xz-$, $yz-$, $yx+$, $zx+$, $zy+$, $yx-$, $zx-$, $zy-\}$,
to benchmark the process tensor prediction. For comparison, we also use the
following traditional quantum map method to predict the quantum process
[Fig.~\ref{fig01}(c)].
\begin{algorithm}[H]
\floatname{algorithm}{Method}
\renewcommand{\thealgorithm}{}
\caption{Quantum Map of the Markovian System}
\begin{algorithmic}[1]
\State Assume that $\mathrm{\mathbf{S}}$ is initially at the ground state.
\State Calculate the state of $\mathrm{\mathbf{S}}$ after the CP map of
$\mathcal{A}_0$ using the $\chi$-matrix determined previously
[Fig.~\ref{fig02}(b)].
\State Calculate the state of $\mathrm{\mathbf{S}}$ after $\mathcal{U}_{1:0}$
using the $\chi$-matrix of the reduced quantum map determined previously.
\label{m1_U}
\State Repeat previous two steps for $\mathcal{A}_1$ and $\mathcal{U}_{2:1}$,
to obtain the final output state.
\end{algorithmic}
Note that in step~\ref{m1_U}, we choose the reduced map of
$\mathcal{U}_{n:n-1}$ conditioned on $\mathrm{\mathbf{E}}$ being in $|0\rangle$.
For example, the $\chi$-matrix on the left panel of Fig.~\ref{fig02}(d).
We assume that $\mathrm{\mathbf{E}}$ remains in $|0\rangle$ throughout
the process, which is the Markov assumption widely used to deal with the open
quantum dynamics but does not always hold.
\end{algorithm}
\label{alg_2}

In the CNOT-CZ process where CNOT gate is activated at the first step and CZ at
the second step, the output states predicted by the process tensor yields
extremely high fidelity and stability, averaged to $99.86\%\pm 1.1\permil$ over
20 repetitions. While the traditional quantum map method with the Markov
assumption can not accurately predict the output of the process,
whose average fidelity is only $80.25\%\pm 13.2\permil$.
A second CZ-CNOT process is also characterized for reference.
Both methods can well predict the process, whose average fidelities are
$99.87\%\pm 1.0\permil$ (process tensor) and
$99.76\%\pm 1.5\permil$ (quantum map of the Markovian system).


\begin{figure}[!htb]
\centering
\includegraphics[width=1.0\columnwidth]{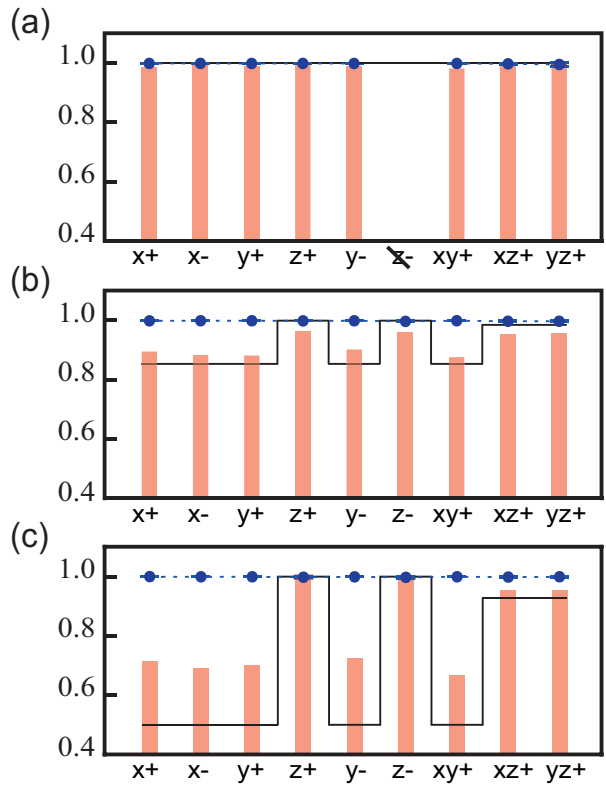}
\caption{
Prediction fidelities of the CNOT-CZ process.
We show three representative groups where qubit is firstly
projected to three states in the $zy$-plane ($\phi=0$),
(a) $\mathcal{A}_0=\mathcal{P}_{z+}$ ($\theta=0$),
(b) $\mathcal{A}_0=\mathcal{P}_{zy+}$ ($\theta=\pi/4$),
(c) $\mathcal{A}_0=\mathcal{P}_{y-}$ ($\theta=\pi/2$).
Prediction fidelities of process tensor method (blue dot) and traditional
quantum map method (solid bar) are drawn on the horizontal axis for different
second-step projections $\mathcal{A}_1$.
Theoretical result of the second method is calculated with all gates ideal
(black line).
Note that one fidelity is omitted when projections are chosen as
$\mathcal{A}_{1:0}$ = $\mathcal{P}_{z+}$-$\mathcal{P}_{z-}$ because
the measurement probability is close to zero.
In other words, this trajectory is forbidden in this process.
}
\label{fig03}
\end{figure}

To study how the previous trajectory of the system affects the subsequent
dynamics, partial results of the CNOT-CZ process are presented in
Fig.~\ref{fig03}. The fidelity of predicted states are grouped by different
choices of $\mathcal{A}_0$:
$\mathcal{P}_{z+}$[Fig.~\ref{fig03}(a)],
$\mathcal{P}_{zy+}$[Fig.~\ref{fig03}(b)],
$\mathcal{P}_{y-}$[Fig.~\ref{fig03}(c)].
We can distinguish the non-Markovian trajectories by checking the discrepancies
of the fidelity.
For example, if we fix the second-step projection to $\mathcal{P}_{x+}$,
the Markov predictions (quantum map method) are unchanged because the
reduced CZ gate is actually an identity map on $\mathrm{\mathbf{S}}$
[left panel of Fig.~\ref{fig02}(d)].
However, the distances between the unchanged state ($\sigma$=$\mathcal{P}_{x+}$)
and the three experimental measured $\rho$ are different,
indicating the change of the final state.
Clearly, the second-step trajectory of $\mathrm{\mathbf{S}}$ depends on the
the first-step operation $\mathcal{A}_0$, a distinguishing feature
of non-Markovianity. In other words, the system dynamics is affected by
the history of operations on it. For the CZ-CNOT process, however, we
can not find any evident changes of the second-step quantum trajectory
conditioned on different $\mathcal{A}_0$.
This means that the CZ-CNOT process is almost Markovian.
Note that the visibility of non-Markovianity in experiment (solid-red bar)
is lower than the ideal one (black reference line).
This is because that both $\mathrm{\mathbf{S}}$ and $\mathrm{\mathbf{E}}$ in our
processor dissipate to a larger lossy environment.
For qubit $\mathrm{\mathbf{S}}$, the measured state and the predicted state
shrink to the north pole of the Bloch sphere, leading to a closer distance
in between.
For qubit $\mathrm{\mathbf{E}}$, it relaxes to the ground state during the
process, losing some memory.

\subsubsection{Quantify the Non-Markovianity}
\label{quantify_NM}

We further quantify the non-Markovianity of the process.
With recorded results of a complete set of POVMs on each step
(Sec.~\ref{sec_ptensors}), we can vary the first-step POVM operator
$\mathcal{P}(\theta,0)$ and derive the rest process tensor,
parameterized as $\mathbf{\mathcal{T}}_{P}^{2:1}(\theta)$.
The azimuth angle $\phi$ is set to zero, i.e. all states of
$\mathrm{\mathbf{S}}$ after the first projections are in the $zy$ plane.

The non-Markovian evolution of a system are held in the process tensor
$\mathbf{\mathcal{T}}_{P}$, which can be conveniently rewritten as a many-body
generalized Choi state $\Upsilon$ using the `Choi-Jamiołkowski'
representation~\cite{ChoiJisomorphism}.
Here we choose to gauge the non-Markovianity by the von Neumann mutual
information~\cite{VedralRMP2002} in $\Upsilon$, which can also be viewed as
the distance between $\Upsilon$ and its uncorrelated state
$\Upsilon_\mathrm{\otimes}$.
The distance is calculated by the von Neumann relative entropy,
\begin{align}
\mathcal{D}(\Upsilon \| \Upsilon_\mathrm{\otimes})=
\operatorname{Tr}
\left[
\Upsilon(
\operatorname{ln} \Upsilon-
\operatorname{ln} \Upsilon_\mathrm{\otimes})
\right].
\label{eq_dist_entropy}
\end{align}

\begin{figure}[!htb]
\centering
\includegraphics[width=1.0\columnwidth]{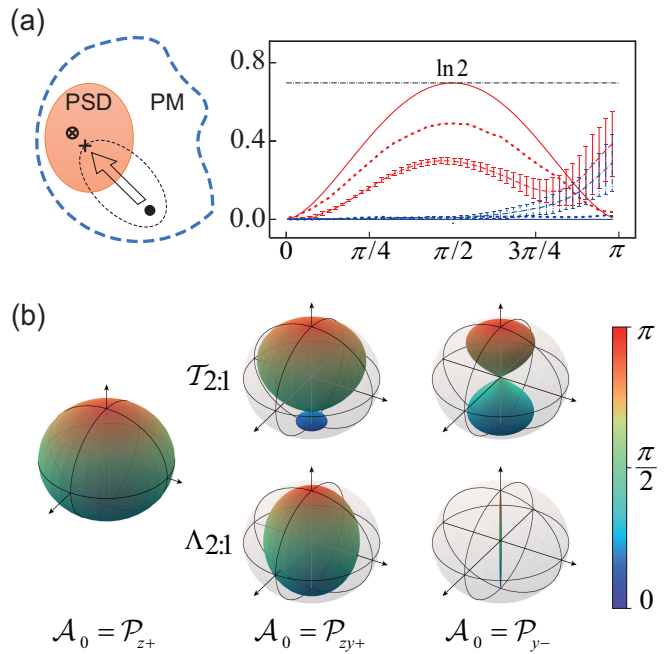}
\caption{
(a) Left panel:
space where the last-two-step restricted process tensor is defined.
The arrow shows the optimization procedure, starting from an initial value
(black dot), to one point $\Upsilon^{2: 1}$ (black cross) that is closet to
the uncorrelated Choi state $\Upsilon^{2: 1}_{\mathrm{\otimes}}$ (cross circle).
Right panel: the non-Markovianity of the last-two-step process
conditioned on different first-step POVMs parameterized with $\theta$.
The CNOT-CZ process yields the maximum non-markovianity when $\theta=\pi/2$.
The experimental result (red dashed line) is lower than the ideal case (red solid line).
Numerical simulations considering dissipation of both $\mathrm{\mathbf{S}}$
and $\mathrm{\mathbf{E}}$ (red dotted line) accounts for the trend.
The other CZ-CNOT process is almost Markovian and memoryless (blue lines).
(b) Volumes of the accessible states of $\mathrm{\mathbf{S}}$.
At the beginning of the second step of CNOT-CZ process,
POVM operators spans the states all around the Bloch sphere (left volume).
When the first-step projection $\mathcal{A}_0$ is $\mathcal{P}_{z+}$,
the final state of the process is unchanged (Identity to $\mathcal{A}_1$),
well predicted by both methods.
When $\mathcal{A}_0$ is $\mathcal{P}_{zy+}$ or $\mathcal{P}_{y-}$,
the process tensor $\mathbf{\mathcal{T}}_{2: 1}$ and the quantum map
$\mathbf{\Lambda}_{2: 1}$ generate the output of the process differently.
We color the states according to the projection angle $\theta$ of
$\mathcal{A}_1$.
}
\label{fig04}
\end{figure}

In our experiment, the Choi state of the last-step process is denoted by
$(\Upsilon^{2: 1})_{4^1\cdot2\times4^1\cdot2}$, with subscripts for its
dimensions.
The uncorrelated state $\Upsilon^{2: 1}_\mathrm{\otimes}$ represents the
last-two-step process with a Markov assumption [Fig.~\ref{fig02}(c)].
We calculate it as the tensor product of the average initial state
of $\mathrm{\mathbf{S}}$ and the Choi state of the reduced quantum map
(conditioned on the average state of $\mathrm{\mathbf{E}}$),
\begin{align}
(\Upsilon^{2: 1}_{\mathrm{\otimes}})_{4\cdot2\times4\cdot2}=
(\Lambda^{2: 1})_{4\times4} \otimes (\rho_{1})_{2\times2}.
\label{eq_prod_state}
\end{align}
Care should be taken when deriving the restricted process tensor
$\mathbf{\mathcal{T}}_{P}^{2:1}$, because it is under-determined using only
POVMs.
The solutions may not always be positive semi-define (PSD).
In contrast, on the set of complete basis including
both projective measurement and unitary control, the full process tensor
is well-determined and PSD~\cite{PollockPRA2018}.
Here we use an optimization method to minimize
the non-Markovianity of $\Upsilon^{2: 1}$ in the space spanned by POVMs
with the PSD constraint.
The procedure to derive $\mathcal{N}^{2: 1}$ is illustrated in the left panel
of Fig.~\ref{fig04}(a).
The restricted process tensor determined by projective measurements
is defined in a linear space spanned by the POVM operators
($\mathrm{\mathbf{PM}}$, dashed-blue line).
It contains a smaller space of process tensors whose Choi states are PSD
(solid-orange area).
In the optimization method, we first find an initial solution
$\Upsilon^{2: 1}_0$ (black dot) using the least-square fitting,
and then progressively optimize the Choi state
$\Upsilon^{2: 1}$ by minimizing the distance [Eq.~\ref{eq_dist_entropy}]
to its uncorrelated state (cross circle $\otimes$) [Eq.~\ref{eq_prod_state}].
We take the minimal distance as the measure of non-Markovianity
$\mathcal{N}^{2: 1}$,
\begin{align}
\mathcal{N}^{2: 1}=
\min_{\Upsilon^{2: 1}\in{\{PSD\}}}
\mathcal{D}(\Upsilon^{2: 1} \| \Upsilon^{2: 1}_\mathrm{\otimes}).
\label{eq_min_psd_entropy}
\end{align}
The result of Eq.~\ref{eq_min_psd_entropy} converges to the point (black cross)
that has the minimal distance to $\Upsilon^{2: 1}_{\mathrm{\otimes}}$.

The right panel of Fig.~\ref{fig04}(b) shows the results of $\mathcal{N}^{2: 1}$
versus different choices of the first-step operator ${\mathcal{P}(\theta, 0)}$.
For the CNOT-CZ process, it yields the maximum non-Markovianity when
$\theta$=$\frac{\pi}{2}$ (red dashed line).
On the contrary, the CZ-CNOT process does not produce significant
non-Markovianity (blue dashed line) whatever $\mathcal{A}_0$ we choose.
As we conclude in Sec.~\ref{sec_ptensors}, the CZ-CNOT process is almost Markovian.

Note that non-Markovianity of the CNOT-CZ process in the real quantum
processor (red dashed line) shows a lowered visibility compared  with
the ideal theoretical result (red solid line).
Similar to the observation in Sec.~\ref{sec_ptensors},
this lowered visibility is due to the dissipation of both
$\mathrm{\mathbf{S}}$ and $\mathrm{\mathbf{E}}$ to a larger environment.
We numerically calculate $\mathbf{\mathcal{T}}_{P}^{2:0}$ using
the non-ideal CNOT and CZ gates that we have experimentally characterized,
and then derive the $\mathcal{N}^{2: 1}$s versus $\theta$ (red dotted line),
from which the trend can be partially verified. The even lower value of
$\mathcal{N}^{2: 1}$ is most likely caused by other noises, such as the
phase noise of a `mediocre' clock when qubit sequence gets
longer~\cite{BarendsNature2014}, and the photon number fluctuations noise
during the projective measurement~\cite{SchusterPRL2005,KrantzAPR2019}.

Both processes show a larger standard deviation when $\theta$ gets bigger.
This is because that we have normalized the Choi state
by dividing the probability of $\mathcal{A}_0$, which reduces to 0
as $\theta$ gets close to $\pi$.
$\mathcal{N}^{2: 1}$s of both processes also get higher when $\theta$
approaches to $\pi$. This indicates that some memory sources are not
included in the simple two-qubit model when $\mathrm{\mathbf{S}}$ is projected
to a higher population of excited-state.
Some possible candidates are the measurement induced state
transitions~\cite{SankPRL2016} and the crosstalk from neighboring qubit.
Nevertheless, such temporal correlation effects will be absorbed in
the process tensor, which can accurately predict the non-Markovian process
(Sec.~\ref{sec_ptensors}).

To illustrate the operational meaning of $\mathcal{N}^{2: 1}$,
we analyze the cause of non-Markovianity in the CNOT-CZ process.
At the beginning of the first process, $\mathrm{\mathbf{E}}$ is reset to
$|0\rangle$.
After one of the first projections
$\mathcal{P}_{z+}$, $\mathcal{P}_{zy+}$, or $\mathcal{P}_{y-}$,
the state of $\mathrm{\mathbf{S}}$ changes to either
$|0\rangle$,
$\mathrm{cos}\frac{\pi}{8}|0\rangle-i\mathrm{sin}\frac{\pi}{8}|1\rangle$,
or $(|0\rangle-i|1\rangle)/{\sqrt{2}}$.
Correspondingly, the state of $\mathrm{\mathbf{SE}}$ changes after the CNOT gate
to either a product state $|00\rangle$, a partial entangled state
$\mathrm{cos}\frac{\pi}{8}|00\rangle-i\mathrm{sin}\frac{\pi}{8}|11\rangle$,
or a maximum entangled state (MES) $(|00\rangle-i|11\rangle)/\sqrt{2}$.
As has been studied in
Refs.~\cite{KuahPRA2007,ModiSRep2012,RingbauerRPL2015,MilzPRA2018},
initial correlations between a system $\mathrm{\mathbf{S}}$ and its
environment $\mathrm{\mathbf{E}}$ carries historical information
and will affect its dynamics at a later time.
At the beginning of the second step, the complete set of POVMs $\mathcal{A}_1$
(after normalization) effectively spans the state all around the Bloch sphere,
[left volume in Fig.~\ref{fig04}(b)].
The process tensor and the quantum map method describe the following process
differently.
We present the volume changes of the set of accessible states of
$\mathrm{\mathbf{S}}$ after the CZ gate in fig.~\ref{fig04}(b).
We first check the quantum map method.
When the $\mathrm{\mathbf{SE}}$ state is $|00\rangle$,
the reduced CZ gate is effectively an identity gate, leaving the states
unchanged (left volume).
When the $\mathrm{\mathbf{SE}}$ state is partially entangled
($\mathrm{cos}\frac{\pi}{8}|00\rangle-i\mathrm{sin}\frac{\pi}{8}|11\rangle$),
some of the phase information of $\mathrm{\mathbf{S}}$ is lost statistically
after the quantum map and the volume of states shrinks toward the $z$-axis
(middle lower volume).
An extreme case is when $\mathrm{\mathbf{SE}}$ is initially in MES.
Phase information of $\mathrm{\mathbf{S}}$ is completely erased through the map,
and all output states goes to the $z$-axis (right lower volume).
Quite differently, the process tensor `knows' how $\mathrm{\mathbf{S}}$ and
$\mathrm{\mathbf{E}}$ are correlated after $\mathcal{A}_1$, which breaks the
entanglement and projects $\mathrm{\mathbf{S}}$ to a predefined state.
Thus the last-two-step process is more accurately described with
tensor operators.
The volume of accessible state predicted by the process tensor
[right upper volumes in Fig.~\ref{fig04}(b)] differs most
with that of the quantum map when $\mathrm{\mathbf{SE}}$ is initially in MES.
This, in turn, is consistent with the non-Markovianity obtained in
Fig.~\ref{fig04}(a). We will most unlikely to confuse the last-two step
process to be Markovian, if we apply $\mathcal{P}_{y-}$ at the first step and
compare the output states with the Markov predictions.
The theoretical value (red solid line) approaching $\ln{2}$ means that the
maximum probability of not finding the process non-Markovian would be
$e^{-\mathcal{N}}=0.5$ for every single ideal experiment.

For the quantification of non-Markovianity, the data analysis code is
available in the repository of Github ~\cite{LiangQucse}.

\section{Conclusion}

In conclusion, we experimentally quantify the non-Markovianity of a quantum
system by intervening projective measurements.
The restricted process tensor is determined with a complete set of POVMs.
Compared with the traditional quantum map method, the process tensor
leads to remarkably high fidelities in predicting the output of an open
quantum process with or without memory effect.
The non-Markovianity of a subset of the process is quantified
conditioned on the choices of the first-step projection.
For the CNOT-CZ process,
we unambiguously determine the exsitance of non-Markovianity and show that
the memory effect is rooted in the spacial correlation between the system and
its environment.
Based on the experiment, we illustrate the operational meaning of the
non-Markovianity: as the non-Markovianity goes high, there is an increased
likelihood to find the Markov assumption wrong.

Although an ancilla qubit is used in our work to simulate the environment,
the process tensor itself is an inclusive model to represent the non-Markovian
noise stemmed from a wide range of
microscopic mechanisms~\cite{MartinisPRB2003,MartinisPRL2005,
BialczakPRL2007,SankPRL2012,SearsPRB2012,RisteNatComm2013,
MalleyPRApp2015,SamachArXiv2021}.
The process tensor method can identify the non-Markovian noise when the
experimenter actively intervenes with the quantum evolution by either
measurement or control.
It will be helpful to analyze and quantify the non-Markovian noise
environment in larger quantum processor~\cite{McEwenNatComm2021}.
Our work also provides a baseline for applying POVMs during the qubits sequence.
Integrated with the process tensor, the measurement based operator can be very
useful in the real-time quantum error correction~\cite{HuNatPhys2019}.
It is also interesting to explore quantum dynamics with varying time steps,
for example, to study the coherent-to-incoherent transition of noise and the
change of memory length.
Since determining a larger tensor network demands exponentially more
resources, we need useful tricks to compress the set of local operators needed.
We can also wisely choose the placement of operations to more efficiently
probe systems of interest.

\begin{acknowledgments}
We thank Kavan Modi for helpful discussion.
The work reported here was supported by the National Key Research and
Development Program of China
(Grant No. 2019YFA0308602, No. 2016YFA0301700),
the National Natural Science Foundation of China
(Grants No. 12074336, No. 11934010, No. 11775129),
the Fundamental Research Funds for the Central Universities in China
(No. 2020XZZX002-01),
and the Anhui Initiative in Quantum Information Technologies
(Grant No. AHY080000).
Y.Y. acknowledge the funding support from Tencent Corporation.
This work was partially conducted at the University of Science and Technology
of China Center for Micro- and Nanoscale Research and Fabrication.

\end{acknowledgments}





\input{biblio_ordered.tex}
\newpage{\pagestyle{empty}\cleardoublepage}

\addcontentsline{toc}{chapter}{\numberline{}\sf\bfseries{Bibliography}}
\end{document}

%% file: biblio_ordered.tex
\newpage{\pagestyle{empty}\cleardoublepage}

%% file: non_markov_tf_refactor.bbl
\begin{thebibliography}{100}

\bibitem{ChuangBook} M. A. Nielsen and I. L. Chuang,
\newblock{Quantum Computation and Quantum Information},
\newblock (Cambridge University Press, Cambridge, England 2010).

\bibitem{Breuer2002OQDbook} H. P. Breuer and F. Petruccione,
\newblock{The theory of open quantum systems},
\newblock (Oxford University Press on Demand, 2002).

\bibitem{Rivas2012OQSBook} 
Á. Rivas and S. F. Huelga,
\newblock{Open Quantum Systems},
\newblock (Springer Berlin Heidelberg, Berlin, 2012).

\bibitem{KossakowskiRMPhys1972} A. Kossakowski,
\newblock{On quantum statistical mechanics of non-Hamiltonian systems},
\newblock Reports Math. Phys. 3, 247 (1972).

\bibitem{GKSJMP1976} V. Gorini, A. Kossakowski, and E. C. G. Sudarshan
\newblock{Completely Positive Dynamical Semigroups of N-Level Systems},
\newblock J. Math. Phys. 17, 821 (1976).

\bibitem{LindbladCMPhys1976} G. Lindblad,
\newblock{On the generators of quantum dynamical semigroups},
\newblock Commun. Math. Phys. 48, 119–130 (1976).

\bibitem{LidarOQS2019} D. A. Lidar,
\newblock{Lecture Notes on the Theory of Open Quantum Systems},
\newblock arXiv:1902.00967 (2019).

%

\bibitem{RivasRPPhys2014} Á. Rivas, S. F. Huelga, and M. B. Plenio,
\newblock{Quantum Non-Markovianity: Characterization, Quantification and Detection},
\newblock Reports Prog. Phys. 77, 094001 (2014).

\bibitem{PollockPRL2018} F. A. Pollock, C. Rodríguez-Rosario, T. Frauenheim, M. Paternostro, and K. Modi,
\newblock{Operational Markov Condition for Quantum Processes},
\newblock Phys. Rev. Lett. 120, 040405 (2018).

\bibitem{MalleyPRApp2015} {P. J. J. O’Malley, J. Kelly, R. Barends, B. Campbell, Y. Chen, Z. Chen, B. Chiaro, A. Dunsworth,
A. G. Fowler, I. C. Hoi, E. Jeffrey, A. Megrant, J. Mutus, C. Neill, C. Quintana, P. Roushan,
D. Sank, A. Vainsencher, J. Wenner, T. C. White, A. N. Korotkov, A. N. Cleland, and J. M. Martinis},
\newblock{Qubit Metrology of Ultralow Phase Noise Using Randomized Benchmarking},
\newblock Phys. Rev. Appl. 3, 044009 (2015).

\bibitem{RingbauerRPL2015} M. Ringbauer, C. J. Wood, K. Modi, A. Gilchrist, A. G. White, and A. Fedrizzi,
\newblock{Characterizing Quantum Dynamics with Initial System-Environment Correlations},
\newblock Phys. Rev. Lett. 114, 090402 (2015).

\bibitem{YuPRL2018} S. Yu, Y. T. Wang, Z. J. Ke, W. Liu, Y. Meng, Z. P. Li, W. H. Zhang, G. Chen, J. S. Tang, C. F. Li, and G. C. Guo,
\newblock{Experimental Investigation of Spectra of Dynamical Maps and Their Relation to Non-Markovianity},
\newblock Phys. Rev. Lett. 120, 60406 (2018).

\bibitem{LuPRL2020} Y.-N. Lu, Y.-R. Zhang, G.-Q. Liu, F. Nori, H. Fan, and X.-Y. Pan,
\newblock{Observing Information Backflow from Controllable Non-Markovian Multichannels in Diamond},
\newblock Phys. Rev. Lett. 124, 210502 (2020).

\bibitem{ChenPRApplied2020} Y.-Q. Chen, K.-L. Ma, Y.-C. Zheng, J. Allcock, S. Zhang, and C.-Y. Hsieh,
\newblock{Non-Markovian Noise Characterization with the Transfer Tensor Method},
\newblock Phys. Rev. Appl. 13, 034045 (2020).

\bibitem{SamachArXiv2021} {G. O. Samach, A. Greene, J. Borregaard, M. Christandl, D. K. Kim, C. M. McNally, A. Melville, B. M. Niedzielski, Y. Sung, D. Rosenberg, M. E. Schwartz, J. L. Yoder, T. P. Orlando, J. I.-J. Wang, S. Gustavsson, M. Kjaergaard, and W. D. Oliver},
\newblock{Lindblad Tomography of a Superconducting Quantum Processor},
\newblock arXiv: 2105.02338 (2021).

\bibitem{BreuerPRL2009} H. P. Breuer, E. M. Laine, and J. Piilo,
\newblock{Measure for the Degree of Non-Markovian Behavior of Quantum Processes in Open Systems},
\newblock Phys. Rev. Lett. 103, 1 (2009).

\bibitem{RivasPRL2010} Á. Rivas, S. F. Huelga, and M. B. Plenio,
\newblock{Entanglement and Non-Markovianity of Quantum Evolutions},
\newblock Phys. Rev. Lett. 105, 1 (2009).

\bibitem{BreuerRevModPhys2016} H. P. Breuer, E.-M. Laine, J. Piilo, and B. Vacchini,
\newblock{Colloquium : Non-Markovian Dynamics in Open Quantum Systems},
\newblock Rev. Mod. Phys. 88, 021002 (2016).

\bibitem{LiPhysRep2018} L. Li, M. J. W. Hall, and H. M. Wiseman,
\newblock{Concepts of Quantum Non-Markovianity: A Hierarchy},
\newblock Phys. Rep. 759, 1 (2018).

\bibitem{LiEPL2019} C.-F. Li, G.-C. Guo, and J. Piilo,
\newblock{Non-Markovian Quantum Dynamics: What Does It Mean?},
\newblock EPL (Europhysics Lett. 127, 50001 (2019).

\bibitem{LiEPL2020} C.-F. Li, G.-C. Guo, and J. Piilo,
\newblock{Non-Markovian Quantum Dynamics: What Is It Good For?},
\newblock EPL (Europhysics Lett. 128, 30001 (2020).

\bibitem{MilzPRL2019} S. Milz, M. S. Kim, F. A. Pollock, and K. Modi,
\newblock{Completely Positive Divisibility Does Not Mean Markovianity},
\newblock Phys. Rev. Lett. 123, 40401 (2019).

\bibitem{CostaNJP2016} F. Costa and S. Shrapnel,
\newblock{Quantum Causal Modelling},
\newblock New J. Phys. 18, 063032 (2016).

\bibitem{MilzQuantum2020} S. Milz, F. Sakuldee, F. A. Pollock, and K. Modi,
\newblock{Kolmogorov Extension Theorem for (Quantum) Causal Modelling and General Probabilistic Theories},
\newblock Quantum 4, 255 (2020).

\bibitem{PollockPRA2018} F. A. Pollock, C. Rodríguez-Rosario, T. Frauenheim, M. Paternostro, and K. Modi,
\newblock{Non-Markovian Quantum Processes: Complete Framework and Efficient Characterization},
\newblock Phys. Rev. A 97, 012127 (2018).

\bibitem{MilzPRA2018} S. Milz, F. A. Pollock, and K. Modi,
\newblock{Reconstructing Non-Markovian Quantum Dynamics with Limited Control},
\newblock Phys. Rev. A 98, 012108 (2018).

\bibitem{XiangPRAppl2020} L. Xiang, Z. Zong, Z. Sun, Z. Zhan, Y. Fei, Z. Dong, C. Run, Z. Jia, P. Duan, J. Wu, Y. Yin, and G. Guo,
\newblock{Simultaneous Feedback and Feedforward Control and Its Application to Realize a Random Walk on the Bloch Sphere in an Xmon-Superconducting-Qubit System},
\newblock Phys. Rev. Appl. 14, 014099 (2020).

\bibitem{WhiteNatComm2020} G. A. L. White, C. D. Hill, F. A. Pollock, L. C. L. Hollenberg, and K. Modi,
\newblock{Demonstration of Non-Markovian Process Characterisation and Control on a Quantum Processor},
\newblock Nat. Commun. 11, 6301 (2020).

\bibitem{PenroseThesis1956} R. Penrose,
\newblock{Tensor Methods in Algebraic Geometry},
\newblock Ph.D. Thesis, Cambridge University (1956).

\bibitem{PenroseCMAppPenrose1971} R. Penrose,
\newblock{Applications of negative dimensional tensors in Combinatorial Mathematics and its Applications, edited by D. Welsh},
\newblock (Academic Press, New York 1971).

\bibitem{BiamonteArXiv2017} J. Biamonte and V. Bergholm,
\newblock{Tensor Networks in a Nutshell},
\newblock arXiv:1708.00006 (2017).

\bibitem{ChiribellaPRL2008} G. Chiribella, G. M. D’Ariano, and P. Perinotti,
\newblock{Quantum Circuit Architecture},
\newblock Phys. Rev. Lett. 101, 1 (2008).

\bibitem{FanoRMP1957} U. Fano,
\newblock{Description of States in Quantum Mechanics by Density Matrix and Operator Techniques},
\newblock Rev. Mod. Phys. 29, 74 (1957).

\bibitem{KuahPRA2007} A. Kuah, K. Modi, C. A. Rodruez-Rosario, and E. C. G. Sudarshan,
\newblock{How State Preparation Can Affect a Quantum Experiment: Quantum Process Tomography for Open Systems},
\newblock Phys. Rev. A 76, 042113 (2007).

\bibitem{VedralRMP2002} V. Vedral,
\newblock{The Role of Relative Entropy in Quantum Information Theory},
\newblock Rev. Mod. Phys. 74, 197 (2002).

\bibitem{ChoiLAA1975} M.-D. Choi,
\newblock{Positive Linear Maps on Complex},
\newblock Linear Algebra Appl. 10, 285 (1975).

\bibitem{ChoiJisomorphism} 
\newblock{Geometry of quantum states: an introduction to quantum entanglement},
\newblock (Cambridge University Press, 2017).

\bibitem{WangPRAppl2019} T. Wang, Z. Zhang, L. Xiang, Z. Jia, P. Duan, Z. Zong, Z. Sun, Z. Dong, J. Wu, Y. Yin, and G. Guo,
\newblock{Experimental Realization of a Fast Controlled-Z Gate via a Shortcut to Adiabaticity},
\newblock Phys. Rev. Appl. 11, 034030 (2019).

%

\bibitem{ZhanArXiv2021} Z. Zhan, C. Run, Z. Zong, L. Xiang, Y. Fei, Z. Sun, Y. Wu, Z. Jia, P. Duan, J. Wu, Y. Yin, and G. Guo,
\newblock{Experimental Determination of Electronic States via Digitized Shortcut-to-Adiabaticity},
\newblock arXiv:2103.06098 (2021).

\bibitem{JeffreySankPRL2014} E. Jeffrey, D. Sank, J. Y. Mutus, T. C. White, J. Kelly, R. Barends, Y. Chen, Z. Chen, B. Chiaro, A. Dunsworth, A. Megrant, P. J. J. O’Malley, C. Neill, P. Roushan, A. Vainsencher, J. Wenner, A. N. Cleland, and J. M. Martinis,
\newblock{Fast Accurate State Measurement with Superconducting Qubits},
\newblock Phys. Rev. Lett. 112, 190504 (2014). 

\bibitem{JohnsonPRL2012} J. E. Johnson, C. Macklin, D. H. Slichter, R. Vijay, E. B. Weingarten, J. Clarke, and I. Siddiqi,
\newblock{Heralded State Preparation in a Superconducting Qubit},
\newblock Phys. Rev. Lett. 109, 050506 (2012). 

\bibitem{MilzOSID2017} S. Milz, F. A. Pollock, and K. Modi,
\newblock{An Introduction to Operational Quantum Dynamics},
\newblock Open Syst. Inf. Dyn. 24, 1740016 (2017).

\bibitem{YamamotoPRB2010} {T. Yamamoto, M. Neeley, E. Lucero, R. C. Bialczak, J. Kelly, M. Lenander, M. Mariantoni,
A. D. O’Connell, D. Sank, H. Wang, M. Weides, J. Wenner, Y. Yin, A. N. Cleland, and J. M. Martinis},
\newblock{Quantum Process Tomography of Two-Qubit Controlled-Z and Controlled-NOT Gates Using Superconducting Phase Qubits},
\newblock Phys. Rev. B 82, 184515 (2010). 

\bibitem{SteffenPRL2006} M. Steffen, M. Ansmann, R. Mcdermott, N. Katz, R. C. Bialczak, E. Lucero, M. Neeley, E. M. Weig, A. N. Cleland, and J. M. Martinis,
\newblock{State Tomography of Capacitively Shunted Phase Qubits with High Fidelity},
\newblock Phys. Rev. Lett. 97, 050502 (2006). 

\bibitem{FlammiaFoundPhys2005} S. T. Flammia, A. Silberfarb, and C. M. Caves,
\newblock{Minimal Informationally Complete Measurements for Pure States},
\newblock Found. Phys. 35, 1985 (2005).

\bibitem{HarrisNumpyNat2020} {C. R. Harris, K. J. Millman, S. J. van der Walt, R. Gommers, P. Virtanen,
D. Cournapeau, E. Wieser, J. Taylor, S. Berg, N. J. Smith, R. Kern, M. Picus,
S. Hoyer, M. H. van Kerkwijk, M. Brett, A. Haldane, J. F. del Río, M. Wiebe,
P. Peterson, P. Gérard-Marchant, K. Sheppard, T. Reddy, W. Weckesser,
H. Abbasi, C. Gohlke, and T. E. Oliphant},
\newblock{Array Programming with NumPy},
\newblock Nature 585, 357 (2020).

\bibitem{JozsaJModOpt1994} R. Jozsa and B. Schumacher,
\newblock{A New Proof of the Quantum Noiseless Coding Theorem},
\newblock J. Mod. Opt. 41, 2343 (1994).

\bibitem{BarendsNature2014} {R. Barends, J. Kelly, A. Megrant, A. Veitia, D. Sank, E. Jeffrey, T. C. White,
J. Mutus, A. G. Fowler, B. Campbell, Y. Chen, Z. Chen, B. Chiaro, A. Dunsworth,
C. Neill, P. O’Malley, P. Roushan, A. Vainsencher, J. Wenner, A. N. Korotkov,
A. N. Cleland, and J. M. Martinis},
\newblock{Superconducting Quantum Circuits at the Surface Code Threshold for Fault Tolerance},
\newblock Nature 508, 500 (2014). 

\bibitem{SchusterPRL2005} {D. I. Schuster, A. Wallraff, A. Blais, L. Frunzio, R.-S. Huang, J. Majer, S. M. Girvin, and R. J. Schoelkopf},
\newblock{Ac Stark Shift and Dephasing of a Superconducting Qubit Strongly Coupled to a Cavity Field},
\newblock Phys. Rev. Lett. 94, 123602 (2005).

\bibitem{KrantzAPR2019} P. Krantz, M. Kjaergaard, F. Yan, T. P. Orlando, S. Gustavsson, and W. D. Oliver,
\newblock{A Quantum Engineer’s Guide to Superconducting Qubits},
\newblock Appl. Phys. Rev. 6, 021318 (2019). 

\bibitem{SankPRL2016} {D. Sank, Z. Chen, M. Khezri, J. Kelly, R. Barends, B. Campbell, Y. Chen, B. Chiaro, A. Dunsworth, A. Fowler, E. Jeffrey, E. Lucero, A. Megrant, J. Mutus, M. Neeley, C. Neill, P. J. J. O’Malley, C. Quintana, P. Roushan, A. Vainsencher, T. White, J. Wenner, A. N. Korotkov, and J. M. Martinis},
\newblock{Measurement-Induced State Transitions in a Superconducting Qubit: Beyond the Rotating Wave Approximation},
\newblock Phys. Rev. Lett. 117, 190503 (2016).

\bibitem{ModiSRep2012} K. Modi,
\newblock{Operational Approach to Open Dynamics and Quantifying Initial Correlations},
\newblock Sci. Rep. 2, 581 (2012).

\bibitem{LiangQucse} L. Xiang,
\newblock{data analysis code as git repository: QUCSE},
\newblock http://github.com/xlelephant/qucse.

\bibitem{MartinisPRB2003} J. M. Martinis, S. Nam, J. Aumentado, K. M. Lang, and C. Urbina,
\newblock{Decoherence of a Superconducting Qubit Due to Bias Noise},
\newblock Phys. Rev. B 67, 094510 (2003).

\bibitem{MartinisPRL2005} J. M. Martinis, K. B. Cooper, R. McDermott, M. Steffen, M. Ansmann, K. D. Osborn, K. Cicak, S. Oh, D. P. Pappas, R. W. Simmonds, and C. C. Yu,
\newblock{Decoherence in Josephson Qubits from Dielectric Loss},
\newblock Phys. Rev. Lett. 95, 210503 (2005).

\bibitem{BialczakPRL2007} R. C. Bialczak, R. McDermott, M. Ansmann, M. Hofheinz, N. Katz, E. Lucero, M. Neeley, A. D. O’Connell, H. Wang, A. N. Cleland, and J. M. Martinis,
\newblock{1/f Flux Noise in Josephson Phase Qubits},
\newblock Phys. Rev. Lett. 99, 187006 (2007).

%

\bibitem{SankPRL2012} {D. Sank, R. Barends, R. C. Bialczak, Y. Chen, J. Kelly, M. Lenander, E. Lucero,
M. Mariantoni, A. Megrant, M. Neeley, P. J. J. O’Malley, A. Vainsencher,
H. Wang, J. Wenner, T. C. White, T. Yamamoto, Y. Yin, A. N. Cleland, and
J. M. Martinis},
\newblock{Flux Noise Probed with Real Time Qubit Tomography in a Josephson Phase Qubit},
\newblock Phys. Rev. Lett. 109, 067001 (2012).

\bibitem{SearsPRB2012} A. P. Sears, A. Petrenko, G. Catelani, L. Sun, H. Paik, G. Kirchmair, L. Frunzio, L. I. Glazman, S. M. Girvin, and R. J. Schoelkopf,
\newblock{Photon Shot Noise Dephasing in the Strong-Dispersive Limit of Circuit QED},
\newblock Phys. Rev. B 86, 180504 (2012). 

\bibitem{RisteNatComm2013} D. Ristè, C. C. Bultink, M. J. Tiggelman, R. N. Schouten, K. W. Lehnert, and L. DiCarlo,
\newblock{Millisecond Charge-Parity Fluctuations and Induced Decoherence in a Superconducting Transmon Qubit},
\newblock Nat. Commun. 4, 1913 (2013). 

\bibitem{McEwenNatComm2021} {M. McEwen, D. Kafri, Z. Chen, J. Atalaya, K. J. Satzinger, C. Quintana, P. V. Klimov, D. Sank,
C. Gidney, A. G. Fowler, F. Arute, K. Arya, B. Buckley, B. Burkett, N. Bushnell, B. Chiaro,
R. Collins, S. Demura, A. Dunsworth, C. Erickson, B. Foxen, M. Giustina, T. Huang, S. Hong,
E. Jeffrey, S. Kim, K. Kechedzhi, F. Kostritsa, P. Laptev, A. Megrant, X. Mi, J. Mutus, O. Naaman,
M. Neeley, C. Neill, M. Niu, A. Paler, N. Redd, P. Roushan, T. C. White, J. Yao, P. Yeh, A. Zalcman,
Y. Chen, V. N. Smelyanskiy, J. M. Martinis, H. Neven, J. Kelly, A. N. Korotkov, A. G. Petukhov, and R. Barends},
\newblock{Removing Leakage-Induced Correlated Errors in Superconducting Quantum Error Correction},
\newblock Nat. Commun. 12, 1761 (2021).

\bibitem{HuNatPhys2019} L. Hu, Y. Ma, W. Cai, X. Mu, Y. Xu, W. Wang, Y. Wu, H. Wang, Y. P. Song, C. L. Zou, S. M. Girvin, L. M. Duan, and L. Sun,
\newblock{Quantum Error Correction and Universal Gate Set Operation on a Binomial Bosonic Logical Qubit},
\newblock Nat. Phys. 15, 503 (2019).

\bibitem{StrathearnNatComm2018} A. Strathearn, P. Kirton, D. Kilda, J. Keeling, and B. W. Lovett,
\newblock{Efficient Non-Markovian Quantum Dynamics Using Time-Evolving Matrix Product Operators},
\newblock Nat. Commun. 9, 3322 (2018).


\end{thebibliography}
